\documentclass{aa}  

\usepackage{natbib,mathtools,graphicx}
\usepackage[varg]{txfonts}

\begin{document}
   \title{Young stellar clusters in the Rosette molecular cloud}
   \subtitle{Arguments against triggered star formation}

   \author{L. Cambr\'esy\inst{1}
	\and
	  	G. Marton\inst{2}
	\and
		O. Feher\inst{3}
	\and
		L.~V. T\'oth\inst{3}
	\and
		N. Schneider\inst{4}
          }

   \institute{Observatoire astronomique de Strasbourg, Universit\'e de
		Strasbourg, CNRS, UMR 7550, 11 rue de l'Universit\'e,
		67000 Strasbourg, France\\
              \email{cambresy@astro.unistra.fr}
	\and
		Konkoly Observatory, Research Centre for Astronomy and Earth
		Sciences, Hungarian Academy of Sciences, Konkoly Thege 15-17,
		H-1121 Budapest, Hungary
	\and
		E\"otv\"os Lor\'and University, Department of Astronomy,
		P\'azm\'any P\'eter s\'et\'any 1/A, 1117 Budapest, Hungary
        \and
		Laboratoire d'Astrophysique de Bordeaux, CNRS/INSU,
		Universit\'e de Bordeaux, BP 89, 33271 Floirac cedex, France
	     }

   \date{Received ; accepted }

\abstract{}
	{We focus on characterizing the young stellar population in the
	\object{Rosette complex} to improve our understanding of the
	processes that regulate the star formation in this region.}
	{We propose an original method that relies on the joint analysis of the
	star color and density in the near--infrared. It leads to mapping
	the molecular cloud spatial distribution and detecting the
	embedded clusters with their characterization in terms of member
	number and age estimation.}
	{We have identified 13 clusters, 2 of which are new discoveries, and we
	estimate that the total number of young stellar objects in the
	Rosette ranges between $4\,000$ and $8\,000$ members.
	We find that the age distribution of the young clusters is not
	consistent with a general triggered scenario for the star formation in
	this molecular cloud.}
	{This study proves that the Rosette complex evolution is not governed by
	the influence of its OB star population. It suggests that the simple
	morphological appearance of an active region is not sufficient
	to conclude much about the triggering role in the star formation
	process.
	Our method of constraining the cluster properties using UKIDSS and
	WISE data has proven efficient, and studies of other regions of
	the galactic plane would definitely benefit from this approach.} 


   \maketitle

\section{Introduction}
The Rosette complex is a well studied region of the galactic plane that
presents the apparent characteristics of a triggered star forming region.
This is, however, still being debated since no strong evidence corroborates
this statement.
It is composed of an OB star cluster that illuminates the famous optical
nebula and interacts with a molecular cloud that contains several star
clusters. The review by \citet{RL08} gives a detailed introduction to this
complex. The well--known region has been observed from the radio to the X-ray
wavelengths. The analysis of the star population in the
\object{Rosette molecular cloud} (RMC) suggests a sequential formation 
\citep{REFL08}, with the more recent clusters farther from \object{NGC~2244}. 

The mechanisms occurring in triggered star forming regions 
\citep[see the review by][]{Elm98} are indeed fully compatible with the
geometry of the Rosette complex, which does appear to be a perfect candidate;
however, the studies based on the interstellar medium rather than the stars
state otherwise. 
\citet{Cel85} observed the radio recombination line in the Rosette H{\sc ii}
and showed the nebula is ionization bound, which suggests that the clusters
outside the ionization front are too far from the central O stars for efficient
triggering to take place. \citet{HWB06} confirm this claim after studying the
turbulent fragmentation in the Rosette. They find that the velocity structures
due to the expansion of ionized gas have not yet propagated through the cloud
to significantly modify its dynamics. \citet{SCH+12} reached the same
conclusion using {\em Herschel} observations to reveal the filamentary structure
of the RMC. They find that clusters lie at the junction of filaments as
predicted by turbulence simulations, including radiative feedback \citep{DB11}.
\citet{PRG+08} obtained Spitzer IRAC and MIPS data and also expressed doubt
about the preponderance of triggering in the Rosette. Their conclusion is
essentially based on the fact that some clusters are outside the ionization
front and that they did not observed any overdensity of very young objects at
the front position itself.
A more local influence has been investigated by \citet{BMR+07}. They
show that the OB stars in NGC~2244 contribute to the photo-evaporation of
a low-mass star circumstellar disk when they are located within 0.5~pc.

In this paper we perform a large scale study of the Rosette complex by
analyzing the interstellar medium distribution and star cluster properties.
The datasets we used are presented in Sect.~\ref{s.data}.
Section~\ref{s.extinction} describes the extinction mapping of the RMC that
allows an unbiased cluster characterization. Section.~\ref{s.ukidss}
focuses on the analysis of the clusters leading to their detection and
the estimation of their total number of members.
The individual identification of members followed by an age study is
presented in Sect.~\ref{s.wise}. We summarize the results and conclude in
Sect.~\ref{s.conclusion}. 
Finally, an appendix provides the reader with a comparison
between the column density derived from background source reddening and from
dust emission in the submillimeter observed with {\em Herschel}.

\section{Datasets}\label{s.data}
The near--infrared $JHK_s$ data were obtained from the UKIDSS \citep{LWA+07}
Data Release 8 (DR8) of the Galactic Plane Survey for a region defined by
the coordinate range $l \in [204.5, 209.0]$ and $b \in [-3.5, -0.5]$.
UKIDSS uses the UKIRT Wide Field Camera \citep[WFCAM;][]{CAA+07}. The
photometric system is described in \citet{HWLH06}, and the calibration is
described in \citet{HIHW09}. The pipeline processing and science archive are
described in \citet{HCC+08}.
The $K_s$ image located at $(l,b)=(206.25,-2.85)$ was not selected for the
DR8 owing its lower quality. The missing sources were taken from
the DR7 source list.
After removing all duplicate sources in UKIDSS, the sample contains
1.4 million objects with a completeness limit of 17.8 mag at $K_s$.
To avoid instrumental saturation issues, the photometry of sources brighter
than 12 mag has been replaced by 2MASS magnitudes at $J$, $H$, and
$K_s$ \citep{SCS+06}.
We also used data from the Wide-field Infrared Survey Explorer
\citep[WISE,][]{WEM+10} for young stellar object (YSO) identification. It
provided us with $2.4\times 10^5$ sources within our field of interest with
the photometry from 3.4 to 22 $\mu$m.

\section{Extinction mapping of the RMC}\label{s.extinction}
The first step in our study consists in mapping the extinction in the whole
molecular cloud. This is required because the dust extinction reddens stars
and decreases their apparent density. Several approaches are possible, such as
using CO or dust emission. \citet{Car00} used CO observations to trace the
extinction before performing a density analysis to extract young clusters,
but CO observations are subject to detection threshold, photodissociation,
depletion, and line saturation. The submillimeter dust emission mapped by
{\em Herschel} is another possibility for tracing the dust distribution
\citep{SMB+10}.
One drawback is the assumption of uniform dust properties, e.g. a
constant emissivity for the whole cloud and a single {\em efficient}
temperature along each line of sight.

The technique we chose for generating the extinction map is described 
extensively in Cambr\'esy et al. (\citeyear{CBJC02,CRMR11}). We only provide
the reader with the outlines of the method here.
Basically, it is an adaptive process where extinction is estimated from the
median color of the three nearest neighbors. The median filters the outliers.
The final map is obtained by a convolution by an adaptive kernel that
produces an extinction map with uniform spatial resolution. This is
different from directly using a regular grid from the start since it minimizes
the nonlinear bias caused by the spatial distribution of the stars,
under-represented at high extinction.

\begin{figure}
\includegraphics[width=8.8cm]{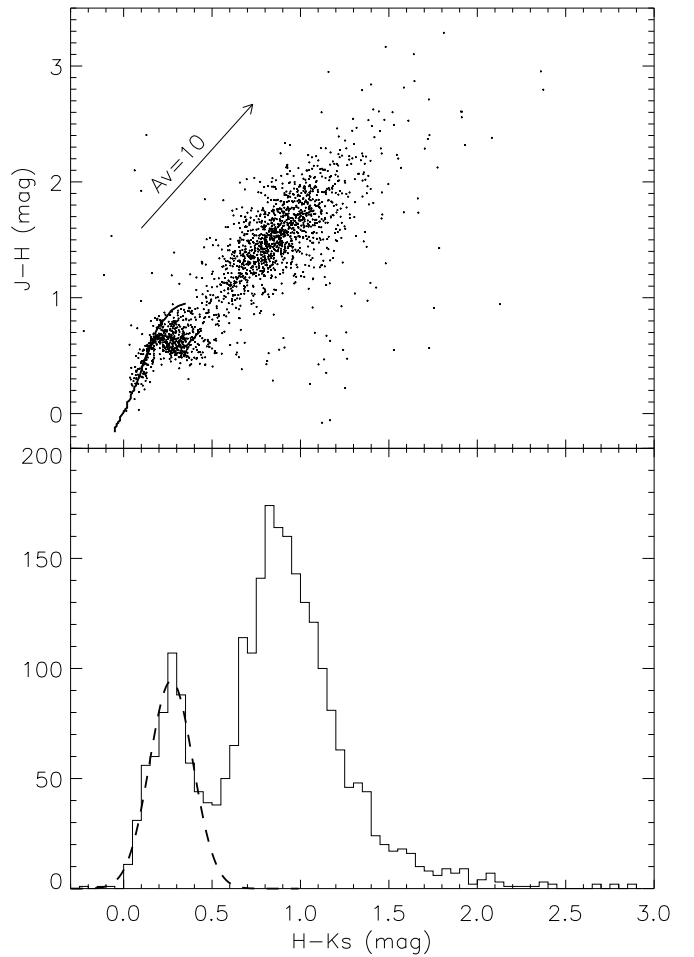}
\caption{{\em Top:} color-color diagram for sources in the regions where
$A_V>10$~mag, excluding the cluster areas. The two groups correspond to
foreground and background stars. The arrow represents the reddening vector
for 10~mag of visual extinction. 
{\em Bottom:} color histogram for the same sources. The dashed line is a
Gaussian fit of the bluer peak, yielding a foreground star density of
$2.2\pm0.1$~arcmin$^{-2}$.}
\label{f.foreground}
\end{figure}

In this section, we restrict our source list to the objects simultaneously
detected at $H$ and $K_s$, with an uncertainty smaller than 0.15~mag and
$K_s<17.8$~mag.
Since the RMC is located in the galactic plane at 1.6~kpc, foreground stars
become the dominant population in the cloud's darkest regions and must be
removed. The degeneracy between intrinsically red foreground stars and blue
background stars prevents their individual classification 
through a color cut-off.
However this degeneracy is raised at high extinction allowing foreground stars
to be identified and their density estimated. Since the foreground object
surface density is uniform over a reasonably large area, an efficient way
to correct from this contamination is to subtract the estimated star density
over the whole field. We found a foreground star surface density of
$2.2\pm0.1$~arcmin$^{-2}$, i.e., $7920\pm340$~deg$^{-2}$ (see
Fig.~\ref{f.foreground}). In the densest part of the cloud, it almost
corresponds to the total star density, which indicates that very few background
objects are actually detected for these lines of sight.
Once the foreground surface density is measured, we still need to decide which
stars to remove. We select the bluer stars, which are the more likely to be
foreground. This truly removes the foreground sources at high extinction.
At lower extinction, the selected stars are not necessarily foreground
because of the degeneracy issue mentioned above. It actually does not 
matter because the total median density over the whole field is
16~arcmin$^{-2}$, which means less than 14\% of the star are foreground.
Removing the bluer ones marginally affects the median color estimation.
At this point, about $10^5$ stars are removed from our source list. It
worth noting that this method has been validated for the more complex case
of the \object{Trifid molecular cloud} \citep{CRMR11}, which is closer to
the Galactic center direction with $(l,b) \approx (7,0)\deg$ and at a larger
distance of 2.7~kpc. A variant but similar approach is proposed by
\citet{KAB+11} for infrared dark clouds.

\begin{figure*}
\includegraphics[width=\textwidth]{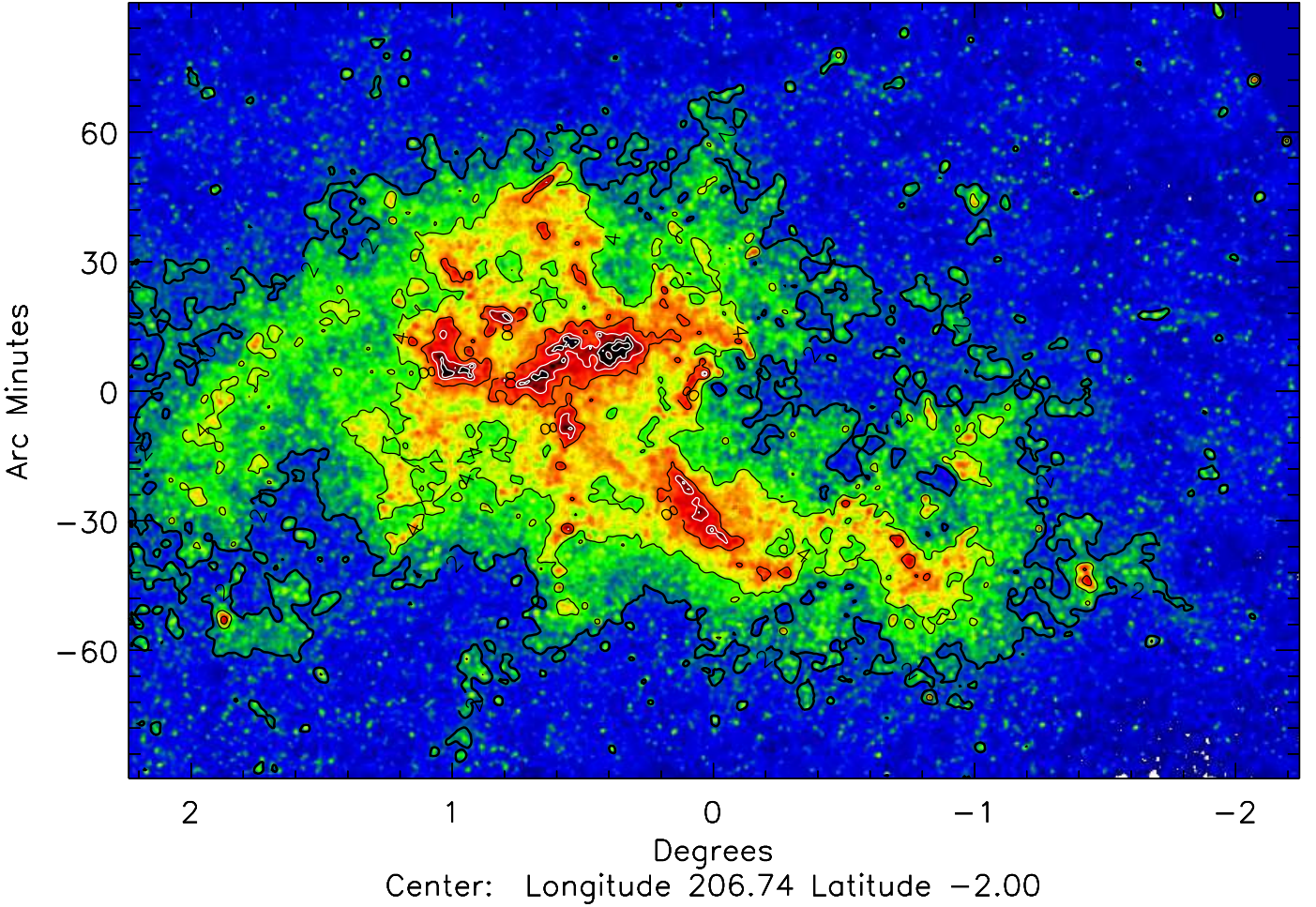}
\caption{Extinction map of the Rosette molecular cloud at 1\arcmin\ resolution.
Contour levels are for visual extinctions of 2, 4, 8~mag in black and 12, 16,
24~mag in white. The maximum extinction peak reaches $A_V=38$~mag.}
\label{f.avRMC}
\end{figure*}

\begin{figure}
\includegraphics[width=8.8cm]{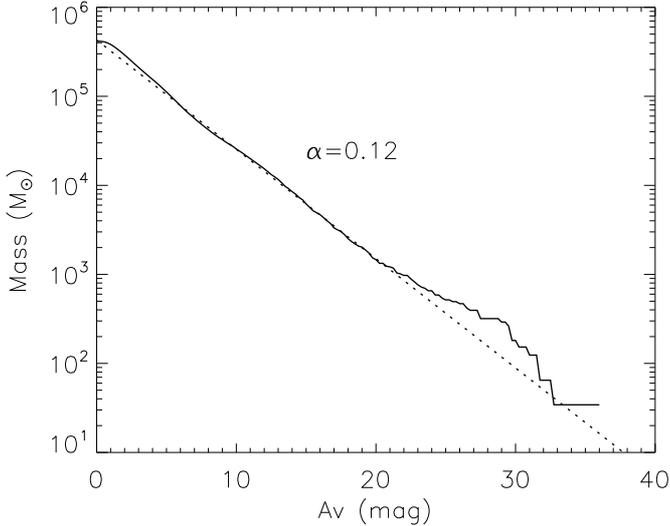}
\caption{Cumulative mass distribution for the Rosette molecular cloud. The
dashed line follows the powerlaw defined by Eq.~\ref{eq.mass}.}
\label{f.mass}
\end{figure}

The final extinction map at a spatial resolution of 1\arcmin\ is presented
in Fig.~\ref{f.avRMC}. It is obtained from $H-K_s$ color excess using the
\citet{RL85} extinction law, which is consistent with our data. The
extinction law variations can be critical in optical wavelength and
beyond 3.5~$\mu$m, but few variations are expected between 1 and 2~$\mu$m.
The extinction is set to zero on the edge of the cloud. No diffuse extinction
along the line of sight has been added so that the extinction map truly
represents the RMC column density.
Assuming a distance of 1.6~kpc and the gas-to-dust
ratio $N_H/A_V=1.87\times 10^{21}$~cm$^{-2}$~mag$^{-1}$ \citep{SM79}, we can
derive the mass of the cloud. The cumulative mass distribution expressed as
the mass enclosed in a given extinction isocontour (see Fig.~\ref{f.mass})
follows this powerlaw for $A_V<20$~mag:
\begin{equation}
M(A_V) = M(0)\times 10^{-\alpha A_V}
\label{eq.mass}
\end{equation}
where the total mass of the cloud $M(0)=4.3\times10^5$~M$_\odot$ and the
index $\alpha=0.122\pm0.005$.
For comparison the total mass of the Orion molecular
cloud is about $3\times 10^5$~M$_\odot$. The relation becomes flatter at
extinctions higher than 20~mag reminding the observed behavior in the Trifid
molecular cloud by \citet{CRMR11}. However, as only 14 arcmin$^2$ reach
this level in the RMC we prefer not to elaborate about this aspect.
Independent mass estimations are available in the literature. 
\citet{WBS95} proposed a mass ranging from $1.1\times10^5$~M$_\odot$
to $2.2\times10^5$~M$_\odot$ from CO observations. This is significantly less
than our value, but mass estimations from CO suffer several issues related to
photodissociation, threshold detection, and line opacity.
\citet{SMB+10} obtained $10^5$~M$_\odot$ from dust emission using
{\em Herschel} data for a smaller surface area. Our mass estimate drops to 
$1.4\times 10^5$~M$_\odot$ if we restrict the initial 13~deg$^2$ of our map to
the 1.8 deg$^2$ of the {\em Herschel} map, making both estimations consistent.

Besides reddening the sources, the extinction produces variations in the star
density. For instance, a visual extinction of 5~mag implies the disappearance
of one third of the background sources at $K_s$. 
The extinction map is essential to correct the stellar density before
performing any clustering analysis, as explained in Sect.~\ref{s.densmap}.

\section{Cluster identification}\label{s.ukidss}
\subsection{Preparation}
\begin{figure}
\includegraphics[width=8.8cm]{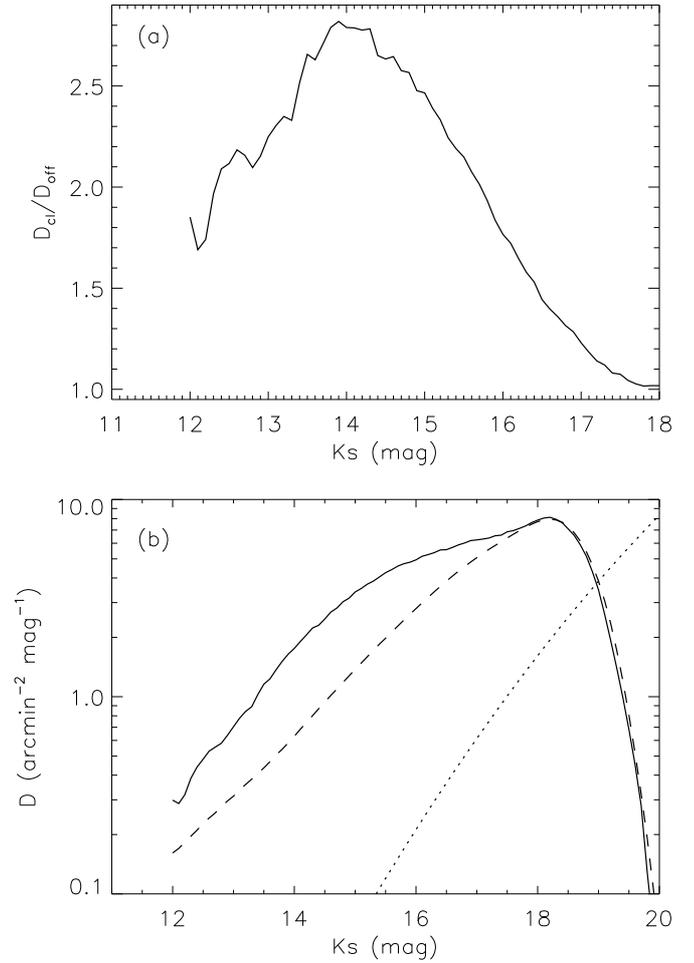}
\caption{{\em Top:} average stellar density ratio for $K_s \in [m-1,m]$
between the cluster's on and off positions.
{\em Bottom:} average stellar density for $K_s \in [m-1,m]$ for the
cluster's on and off positions with solid and dashed lines, respectively. The
dotted line represents the density of galaxies \citep{GSFC97}.}
 \label{f.contrast}
\end{figure}

A quick-look surface density analysis is obtained through a 1\arcmin\ sampling
star count of the $K_s$ sources. Instead of a map we actually built a datacube
where each plane is a density map for sources up to a magnitude limit ranging
from 11 to 20 mag in steps of 0.1 mag. The main star clusters are detectable by
eye when browsing the datacube from the bright to faint stars.
This datacube permits the magnitude distribution of the cluster members to
be probed using the differential stellar density, defined as the density of
stars with $K_s \in [m-1,m]$. The differential stellar density ratio between 
pixels covering the cluster areas ({\em on}) and pixels excluding the
cluster areas ({\em off}) increases with the magnitude up to about
$K_s \approx 14$~mag and then decreases when counting fainter sources
(Fig.~\ref{f.contrast}a). 
Figure~\ref{f.contrast}b presents the two differential densities for the
cluster on and off positions.
The plot indicates that the clusters essentially contain stars brighter than
17~mag at $K_s$, which is a direct consequence of their distance and age
population. These plots teach us that the signal-to-noise ratio in the
clustering analysis will be enhanced by restricting the sample to sources
with $K_s \in[12.5,15.5]$~mag. This range is a good compromise with a high
density contrast and a sample that still includes about 65\% of the
cluster members. Finally, the question of the contamination by galaxies
must be addressed. It appears to be negligible for this magnitude limit
with an estimated density of $\sim 0.1$~arcmin$^{-2}$, about 50 times less
than the star density as shown Fig.~\ref{f.contrast}b.

\subsubsection{Density map}\label{s.densmap}
\begin{figure}
\includegraphics[width=8.8cm]{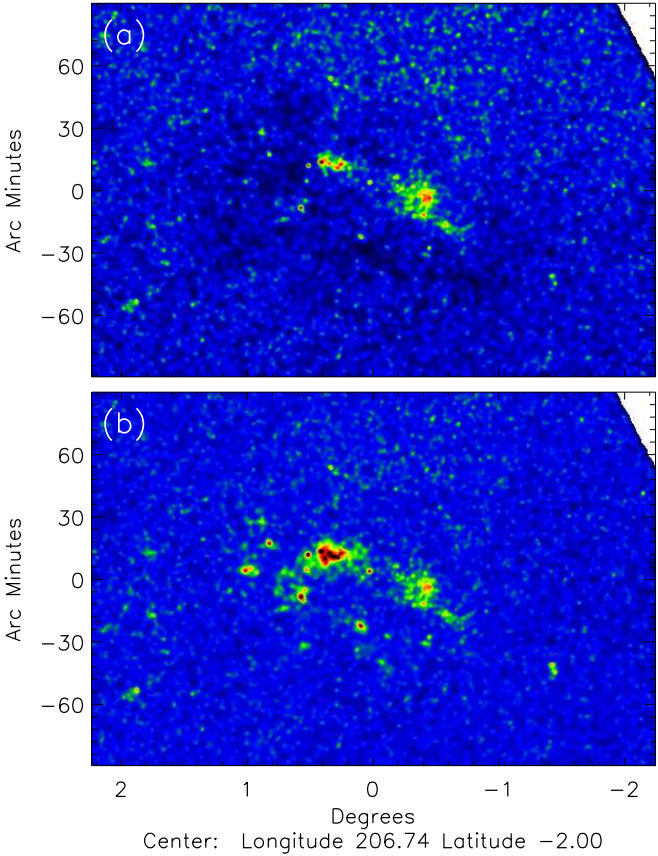}
\caption{{\em Top:} raw stellar density map. {\em Bottom:} stellar density map
corrected for the interstellar extinction.}
 \label{f.density}
\end{figure}

The historical and trivial technique for mapping the stellar surface density
is to count stars within each cell of a regular grid. 
Such a density estimator is actually the classical 2D histogram for which cell
size has to be defined in advance for the whole field, and the grid initial
position matters. 
The concept of an adaptive grid used for the extinction mapping yields an
improved density estimator. It consists of setting the number of sources
per cell rather than the grid size. The local density, $D$, is then
derived from the distance, $r$, of the $N^{th}$ nearest neighbor by
\begin{equation}
D=\frac{N-1}{\pi r^2}.
\label{eq.density}
\end{equation}
\citet{CH85} demonstrated that counting the $N^{th}$ star produces a
bias in the density estimation. This is why $N-1$ is preferred in the
Eq.~\ref{eq.density} numerator. There are other statistical methods for
estimating the density \citep[see for example][]{Sil86}, but the 
nearest-neighbor method
has proven to be efficient for stellar distributions and is indeed
widely used in the Rosette complex \citep{REFL08,PRG+08,WFT+09}. By using
$N=20$ for stars with $K_s \in [12.5,15.5]$~mag, this method leads to the
density map presented in Fig.~\ref{f.density}a.
Its median stellar density is $3.1\pm0.5$~arcmin$^{-2}$. 
We define clusters as a density threshold 3$\sigma$ above that level, 
i.e., above 4.6~arcmin$^{-2}$.
The corresponding spatial resolution for the mapping is 1.1\arcmin\ on the
cluster edge, it is better inside since the stellar density is higher. 

The source density (Fig.~\ref{f.density}a) and the extinction maps 
(Fig.~\ref{f.avRMC}) share anticorrelated structures. This is the expected
consequence of extinction that makes the stellar density decrease.
The extinction structures, which trace the interstellar matter distribution,
are background variations of the star surface density map. They need to be
subtracted in order to apply a cluster detection algorithm on a flat density
map. Once the extinction map has been built, these background variations are
easily predictable and the corrected map is obtained as follows:
\begin{equation}
D_\mathrm{corr}=D\times 10^{\,a\,A_{K_s}}
\label{eq.corr}
\end{equation}
where $A_{K_s}=0.112\, A_V$, $a=0.318$ is our measure of the slope
of the $K_s$ luminosity function, and $D$ and $D_\mathrm{corr}$ are the
observed and the true stellar densities, respectively. The true,
absorption--free, density map is presented in Fig.~\ref{f.density}b.
The molecular cloud structures are no longer visible. The background is
flat and set to zero by subtracting the median value of the map. The result
is a field-star--subtracted surface density map for which structures are true
stellar overdensities.

It is important to realize that all stars have been implicitly assumed to be
background when using Eq.~\ref{eq.corr}, whereas this is unknown for the
cluster members themselves. Clusters can be either embedded or on the rear
or the front side of the molecular cloud, and they will not suffer from
the same amount of extinction depending on the case. Once the correction
is done, setting the background to zero subtracts the field star density
everywhere, including toward clusters. It is then possible to reverse the
extinction correction on the clusters only assuming they would be on the
front side rather than the rear side of the cloud. The final result for the 
overdensity map $D_0=D_\mathrm{corr}-\overline{D}_\mathrm{corr}$, with 
$\overline{D}_\mathrm{corr}$ the median density, can be written as
\begin{equation}
D_0 = \underbracket[0.7pt][10pt]{ D^\mathrm{field}\times 10^{\,a\,A_{K_s}} - 
			\overline{D}_\mathrm{corr} }_\text{noise, $\sigma=0.5$~arcmin$^{-2}$} +
      \left\{\begin{array}{ll}
        D^\mathrm{cl}\times  10^{\,a\,A_{K_s}} & \text{back cluster}\\
	\\
        D^\mathrm{cl}                          & \text{front cluster}
      \end{array}\right.
\label{eq.corr2}
\end{equation}
where $D^\mathrm{field}$ and $D^\mathrm{cl}$ stand for field star and
cluster densities, respectively.
Both cases are taken into account when estimating the number of
members per cluster in the next section.

\subsubsection{Cluster characterization}
\begin{figure*}
\includegraphics[width=\textwidth]{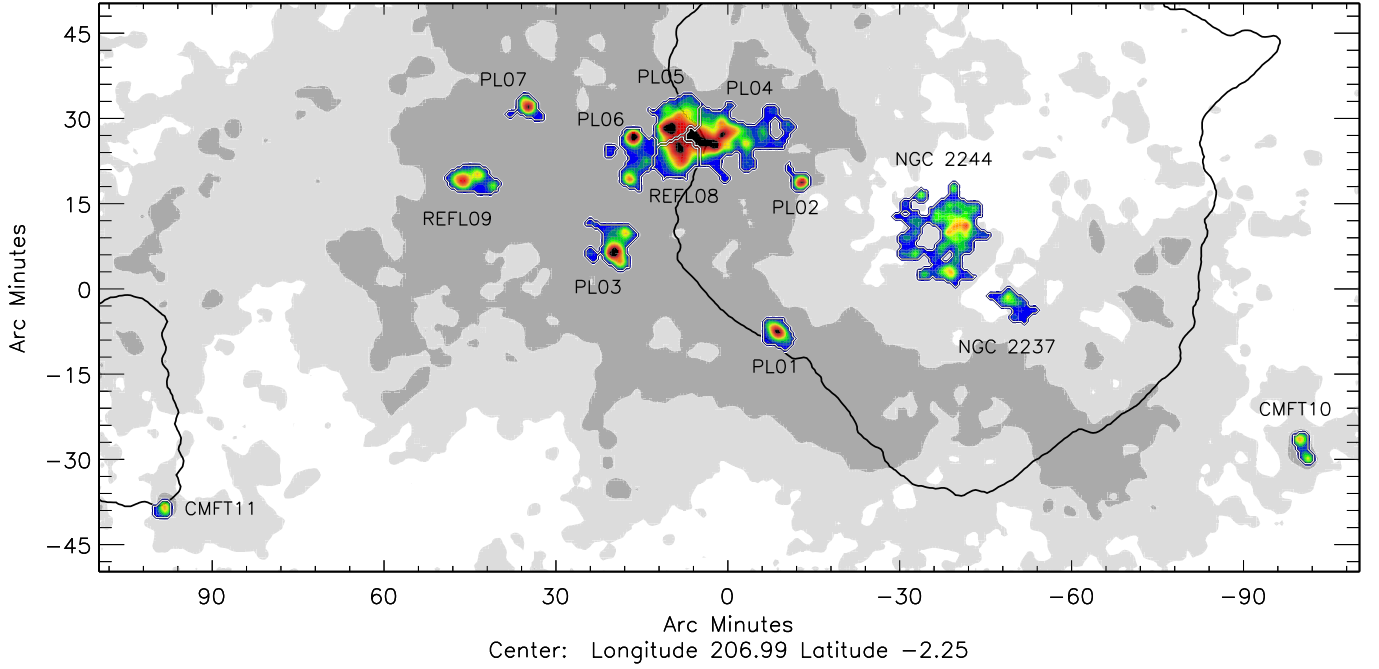}
\caption{Cluster detections in the Rosette molecular cloud. PL and REFL
designations are from \citet{PL97} and \citet{REFL08}, respectively.
The solid black line shows the H$\alpha$ contour that indicates the
ionization front of the Rosette H{\sc ii} nebula centered on NGC 2244 and
of \object{Sh~2-280} to the east of the map. The two levels of shaded gray
correspond to visual extinction of 2 and 4~mag, respectively. CMFT~10
and CMFT~11 are new clusters.}
 \label{f.clusters}
\end{figure*}

We applied the clump detection algorithm {\em clumpfind} developed by
\citet{WGB94} on our density map. Its original 3D implementation for
spectral line datacubes was generalized to 2D for submillimeter maps.
Although {\em clumpfind} was designed to detect gas or dust clumps, it
finds structures by connecting pixels of similar values, regardless of what
the data represent. The detected structures in the star density map are
stellar clusters by definition.
In the {\em clumpfind} philosophy, each peak is a clump maximum. The
peaks are selected above a level provided by the user, and the lowest
level is the threshold for detection. Obviously the threshold definition would
have been a serious issue with the raw density map with no extinction
correction, since its background level is not uniform. We applied the IDL
version of the clumpfind algorithm on the corrected density map
(Fig.~\ref{f.density}b) with the levels 3, 20, 25, and 40~arcmin$^{-2}$,
meaning a cluster must have a star density excess greater than 3~arcmin$^{-2}$. 
Figure~\ref{f.clusters} shows the result with the clusters identification
from \citet{PL97} and \citet{REFL08}.
\begin{table*}
\caption{Cluster characteristics. Column (2) contains the cluster geometrical
centers followed by their dimensions. Column (4) is positions of the density
peaks followed by two estimations for the peak value assuming a back or
a front side location of the clusters. Columns (6) and (7) are the
statistical numbers of members within clusters for back and front
clusters; column (8) provides the total number of UKIDSS sources
used to derive the extinction ($K_s<17.8$~mag), and column (10) gives
the average visual extinction for the cluster.}\label{t.stat}
\begin{tabular}{lccccccrrrrr}
\hline
\hline
Name & Glon \: Glat & $\Delta$Glon & $\Delta$Glat & Glon$^\mathrm{peak}$ \:
Glat$^\mathrm{peak}$ &  Max$^\mathrm{B}$ & Max$^\mathrm{F}$ & N$^\mathrm{B}$& N$^\mathrm{F}$& N$^\mathrm{cat}$ & Area & $\overline{A_V}$ \\
     & deg & \multicolumn{2}{c}{arcmin} & deg & \multicolumn{2}{c}{arcmin$^{-2}$} & & &{\scriptsize UKIDSS} & arcmin$^2$ & mag \\
(1)  & (2) & \multicolumn{2}{c}{(3)}    & (4) & \multicolumn{2}{c}{(5)} &(6)&(7)&(8)&(9)&(10) \\
\hline
NGC2244& 206.368\: -2.096&17.5&20.0& 206.293\: -2.067& 17.2& 14.3& 912&763 &3840 &169.7& 2.2\\
NGC2237& 206.160\: -2.304& 8.8& 7.1& 206.176\: -2.275&  9.6&  7.9& 132&106 & 668 & 29.7& 2.6\\
PL01   & 206.839\: -2.387& 5.6& 5.6& 206.843\: -2.383& 40.3& 11.9& 203& 77 & 387 & 26.5&10.9\\
PL02   & 206.785\: -1.925& 4.3& 4.6& 206.777\: -1.942& 29.8&  9.5& 102& 46 & 195 & 13.5& 8.5\\
PL03   & 207.327\: -2.125& 7.8& 9.3& 207.319\: -2.150& 79.8& 23.2& 433&174 & 875 & 51.7& 9.6\\
PL04   & 206.952\: -1.825&20.1&10.5& 207.027\: -1.833& 53.9& 17.0&1241&536 &2461 &136.7& 8.6\\
PL05   & 207.143\: -1.758& 8.5& 8.6& 207.152\: -1.783& 87.5& 32.6& 683&293 &1096 & 55.7& 9.2\\
REFL08 & 207.144\: -1.871& 7.7& 7.8& 207.127\: -1.842& 39.4&  5.5& 533&125 & 473 & 46.0&16.0\\
PL06   & 207.273\: -1.875& 7.3&14.1& 207.260\: -1.808& 73.3& 18.6& 394&137 & 670 & 57.7&12.1\\
PL07   & 207.581\: -1.721& 5.5& 4.7& 207.569\: -1.716& 28.7&  9.3& 140& 53 & 238 & 19.2&10.8\\
REFL09 & 207.731\: -1.933&10.4& 5.5& 207.760\: -1.941& 20.8&  5.3& 265& 88 & 463 & 39.7&12.7\\
CMFT10   & 205.313\: -2.720& 4.2& 7.7& 205.325\: -2.691& 12.2&  6.5&  91& 52 & 305 & 15.5& 6.5\\
CMFT11   & 208.636\: -2.898& 3.8& 3.8& 208.628\: -2.890& 12.4&  6.4&  60& 38 & 212 & 10.2& 5.0\\
\hline
\end{tabular}
\end{table*}
The levels 20, 25, and 40~arcmin$^{-2}$ are arbitrary chosen to split the
connected clusters \object{PL~04}, \object{PL~05}, \object{PL~06}, and 
\object{REFL~08} in accordance with \citet{REFL08}. As pointed out by
\citet{PRG09}, the {\em clumpfind} decomposition in {\em crowded} regions
does not provide a robust list of physically meaningful clumps, although
it is relevant for statistical analysis. We found 13 clusters 
(NGC~2244, \object{NGC~2237}, PL~01--07 from \citealt{PL97}; REFL~08,09
from \citealt{REFL08}) and two new clusters \object{CMFT~10} and
\object{CMFT~11}, which we named following the other cluster designation
in this region using the author's initials; we do not confirm the existence
of \object{REFL~10}.

The total number of members for each cluster is obtained by
integrating the density map over their surface area; however, the extinction
correction applied with Eq.~\ref{eq.corr} on the whole map is only justified
for clusters on the back of the molecular cloud and is irrelevant
if they are located on the front side. As mentioned above, it is
possible to compute the member numbers assuming cluster are on the front
by reversing the correction in agreement with Eq.~\ref{eq.corr2}.
The results presented in Table~\ref{t.stat} provide the statistical number
of members assuming the cluster is on the back or the front side of
the cloud. The more the extinction towards a cluster, the greater the
difference between these two numbers. The most affected cluster is
REFL~08 with an estimated population that varies by a factor of 4
depending its location along the line of sight.
The number of YSOs in Table~\ref{t.stat} are for $K_s<15.5$~mag,
which corresponds to 65\% of the total star overdensity following
Fig.~\ref{f.contrast}. Therefore the total number of YSOs in the Rosette
complex is expected to range between 4\,000 and 8\,000 members.
We stress that this estimate misses only a small fraction of the total
number of YSOs, which are the very young protostars (Class 0) that are
undetectable at $K_s$ because they emit mostly in the far--infrared and 
submillimeter wavelengths. For instance, the magnitude of a Class I source
of $10^6$~years and only 0.1~M$_\odot$ would be $K_s\approx 15.5$~mag
according to the model tool described in \citet{SDF00}. This is easily
detected with UKIDSS, and it explains why the stellar surface density at
$K_s>17$~mag does not exhibit any more excess (see Fig.~\ref{f.contrast}).
Deeper observations at this wavelength would not improve YSO detection rate. 

The number of YSOs permits the star formation efficiency (SFE) to be estimated
per cluster. It is defined as the ratio of the YSO mass over the total
mass (YSOs and gas). The YSO mass is derived from their number given in
Table~\ref{t.stat} for the back and the front cases (columns 6 and 7)
and assuming a mean stellar mass of 0.6~M$_\odot$. The gas mass is directly
measured on the extinction map within each cluster boundary.
Table~\ref{t.sfe} presents the results. For NGC~2244 and NGC~2237 the SFE
is likely overestimated since these two clusters are located in the region
where the gas has been swept up by the OB stars. For the others,
the values range between 10 to 24\% or 3 to 12\% depending the assumed 
location of the clusters along the line of sight. This is to compare with
the maximum efficiency of $\sim 30\%$ derived from the mass function of dense
molecular cores \citep{ALL07}. The observed SFE actually depends on the
scale, from 10--30\% for embedded clusters to only 1--5\% for entire giant
molecular clouds \citep{LL03}. This suggests that the RMC activity has not
come to an end. PL~05 and CMFT~11 appear to be the most efficient at forming
stars.

\begin{table}
\caption{Star formation efficiency for each cluster. Two SFE values are
proposed to account for the uncertainty on the cluster locations with respect
to the molecular cloud (the superscript B and F letters refer to the back
and front side location as in Table~\ref{t.stat}).}\label{t.sfe}
\begin{tabular}{lrcc}
\hline
\hline
Name & Gas mass      & SFE$^{\rm B}$ & SFE$^{\rm F}$ \\
     & M$_\odot$\hspace{1.3ex} & \%            & \%            \\
\hline
NGC 2244 & 1518\hspace{1ex} & 36.1 & 30.2 \\
NGC 2237 &  316\hspace{1ex} & 28.1 & 22.6 \\
PL01     & 1201\hspace{1ex} & 13.7 &  5.2 \\
PL02     &  469\hspace{1ex} & 17.0 &  7.6 \\
PL03     & 2075\hspace{1ex} & 16.4 &  6.6 \\
PL04     & 4834\hspace{1ex} & 19.4 &  8.4 \\
PL05     & 2049\hspace{1ex} & 23.8 & 10.2 \\
REFL08   & 3133\hspace{1ex} & 13.8 &  3.2 \\
PL06     & 2893\hspace{1ex} & 11.3 &  3.9 \\
PL07     &  901\hspace{1ex} & 12.7 &  4.8 \\
REFL09   & 2163\hspace{1ex} & 10.3 &  3.4 \\
CMFT10   &  472\hspace{1ex} & 15.3 &  8.7 \\
CMFT11   &  233\hspace{1ex} & 19.5 & 12.3 \\
\hline
\end{tabular}
\end{table}

\section{Individual members}\label{s.wise}
This section focuses on the study of the 13 identified and well-delimited
clusters to explore the role of triggering by the NGC~2244 OB stars in more
detail.
The UKIDSS deep near--infrared data can be associated with the WISE
observations at longer wavelengths to classify the sources located within
the projected cluster boundaries.
The classification process is complex because of the multiple contaminants,
which are extragalactic sources, shock emission blobs, or resolved PAH
emission structures. \citet{KLB+12} propose a method of filtering all these
sources and extracting YSOs. We started with their recommendations for
building a first sample, and we improved completeness using a second method
based on quadratic discriminant analysis. It is a statistical method for
classification and pattern recognition in multidimensional data by looking
for quadratic boundaries between groups of different object types.
The training sample was prepared with the use of objects from previously
known types, i.e. galaxies, main-sequence stars, evolved stars, and YSOs,
all listed in SIMBAD and VizieR. For a detailed description of the selection
scheme see Marton et al. (2013, in preparation).
\begin{figure*}
\includegraphics[width=\textwidth]{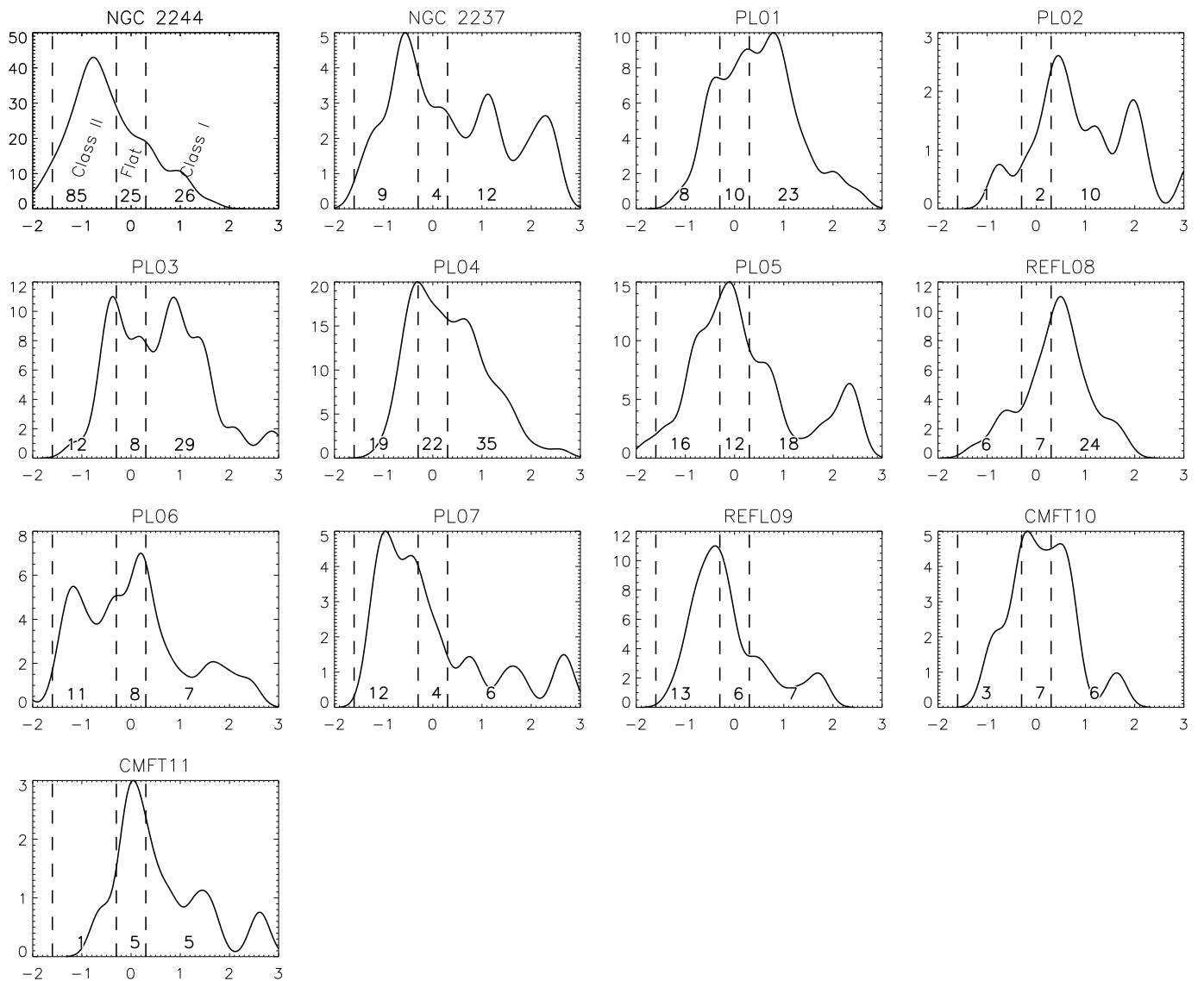}
\caption{Density distribution of the spectral index 
$\alpha=d\log{(\lambda F_\lambda)}/d\log{\lambda}$.
The dashed lines define the Class II, Flat spectrum, and Class I source
locations. The number of objects for these three classes is indicated
for each cluster. 
The density distribution is obtained using the kernel estimator 
presented in \citet{Sil86}. It is particularly well adapted to small
samples since it does not depend on the first bin position and is
oversampled.}
 \label{f.alpha}
\end{figure*}

We classified the YSOs according to the slope of the spectral index
$\alpha = d\log{(\lambda F_\lambda)}/d\log{\lambda}$ \citep{GWA+94}. The
source list of 535 YSO candidates with astrometry, photometry, and spectral
index is published electronically at the CDS. Table~\ref{t.yso} is a
subsample that includes all candidates within the two new clusters CMFT~10
and CMFT~11.
The spectral index is derived from a least square fitting from the $K_s$
band to the third WISE channel, $W3$, at 12~$\mu$m, whose passband actually
extends from 8 to 18~$\mu$m. Sources are classified as Class I, flat spectrum,
and Class II YSOs. Figure~\ref{f.alpha} shows the spectral index histogram
for all clusters and the number of sources in each class. The spectral index
depends mainly on the disk size and orientation, which implies its conversion
into source age is not direct. However, it can be tested statistically
with the spectral energy distribution (SED) fitting tool proposed by
\citet{RWIW07}, which provides the 10\,000 best possible solutions for a
given energy distribution. We examined the correlation of $\alpha$ with the
1\,000 best fits for each star in two clusters, PL~07 and CMFT~11, which allow
probing of a wide range of values. While $\alpha$ is measured only with the
four fluxes from $K_s$ to $W3$, the SED fitting is performed on the seven
fluxes from $J$ to $W4$ to better constrain the source physical properties.
The correlation between $\alpha$ and several parameters, such as the age,
the central source mass, the disk inclination, the accretion rate, or the
extinction, is analyzed by measuring the histogram peak of the 1\,000 best
fits for each of these parameters. The only parameters that exhibits a
clear correlation (coefficient of 0.9) with the spectral index is the
age that follows the powerlaw
\begin{equation}
\log{(\rm Age-10^4}) = (-0.88\pm 0.07) \alpha + (5.02\pm 0.08).
\label{eq.alpha}
\end{equation}
Since the age provided by the \citet{RWIW07} model starts with the theoretical
evolutionary tracks; i.e., after the protostar collapse, we arbitrarily added
$10^4$ years to refer more specifically to an earlier stage.
The age obtained from the fitting model is actually indirectly determined
through constraining the circumstellar parameters and is used to give a rough
idea of the stellar properties as a function of mass (T. Robitaille, private
communication).  It cannot be considered as a reliable age measurement for
individual stars, so we carefully limit its usage to a statistical analysis. 
Under this restriction, it is possible to interpret Fig.~\ref{f.alpha} as
an age distribution.
This histogram representation is a valuable tool that provides us with
an immediate and comprehensive picture of the age distribution within a
cluster. 
Table~\ref{t.wise} provides the median age estimate for
each cluster, together with the 10\% and 90\% quantiles of the sample.
Quantiles are more robust to statistical fluctuations and yield an estimation
of the cluster ages, which we believe is correct within a factor of 2.
The 10\% quantile represents the age of the young population of
the clusters, whereas the 90\% quantile provides us with the age of the
oldest population and therefore indicates the age of the beginning of
the star formation in each cluster.
\begin{table}
\caption{Age estimation of the cluster members detected by WISE based on
their spectral index using Eq.~\ref{eq.alpha}. The columns labeled
{\em young} and {\em old} contain the age estimate for the youngest and
the oldest populations within each cluster.}\label{t.wise}
\begin{tabular}{lcccc}
\hline
\hline
Name   & Number & \multicolumn{3}{c}{Age (yr)}\\
       &        & young & median & old \\
\hline
NGC 2244  & 145 & $3.6\times 10^4$ & $3.8\times 10^5$ & $1.8\times 10^6$ \\
NGC 2237  &  25 & $1.1\times 10^4$ & $8.7\times 10^4$ & $1.0\times 10^6$ \\
PL01      &  41 & $1.5\times 10^4$ & $5.8\times 10^4$ & $2.9\times 10^5$ \\
PL02      &  13 & $1.1\times 10^4$ & $3.3\times 10^4$ & $2.0\times 10^5$ \\
PL03      &  49 & $1.1\times 10^4$ & $3.3\times 10^4$ & $2.7\times 10^5$ \\
PL04      &  76 & $1.6\times 10^4$ & $8.0\times 10^4$ & $3.2\times 10^5$ \\
PL05      &  47 & $1.1\times 10^4$ & $1.2\times 10^5$ & $6.1\times 10^5$ \\
REFL08    &  37 & $1.8\times 10^4$ & $5.2\times 10^4$ & $3.7\times 10^5$ \\
PL06      &  27 & $1.3\times 10^4$ & $1.1\times 10^5$ & $1.5\times 10^6$ \\
PL07      &  22 & $1.2\times 10^4$ & $2.4\times 10^5$ & $1.1\times 10^6$ \\
REFL09    &  26 & $1.9\times 10^4$ & $2.1\times 10^5$ & $5.8\times 10^5$ \\
CMFT10    &  16 & $3.4\times 10^4$ & $9.4\times 10^4$ & $5.2\times 10^5$ \\
CMFT11    &  11 & $1.2\times 10^4$ & $7.2\times 10^4$ & $2.1\times 10^5$ \\
\hline
\end{tabular}\\
\footnotesize{{\em young} and {\em old} refer to the 10\% and 90\% quantile,
respectively.}
\end{table}

\noindent
The oldest cluster is NGC~2244, which starts forming stars about 2~Myr,
in agreement with
the usual age determination of its O star population. Then the clusters
tend to be younger as the distance to NGC~2244 increases. This is true
for NGC~2237, \object{PL~01}, \object{PL~02}. The youth of PL~01 and
PL~02 would be consistent with a scenario where compression effects can
indeed locally provoke the formation of clusters in the immediate
interaction zone between the H{\sc ii}-region and molecular cloud
(on a parsec scale of a few). \citet{SCH+12}  arrive at the same
conclusion based on the large number of dense cores found in this
interaction region. The age of \object{PL~03} is similar, but it is
located beyond the influence of NGC~2244. PL~05 is even older than these
potentially triggered clusters, with about $6.1\times10^5$~yr.
Going farther, PL~06 and \object{PL~07} are older at more than 1~Myr
and may even have formed at the same time as NGC~2244.
\citet{SMB+10} and \citet{HMB+10} argue that PL~07 would be the youngest
cluster in the RMC, but their age classification was determined in a
preliminary way and with lower quality statistics. We agree that there are
young protostars in this cluster, but the relevant age that determines the
beginning of the collapse is the one of the oldest population of the cluster,
not the youngest.
The collapse of the more distant clusters, \object{REFL~09}, CMFT~10, and
CMFT~11, are similar to PL~04 whereas they are definitely out of reach of
the OB stars in NGC~2244. We cannot rule out the possibility that some
cluster formation has been triggered by NGC~2244, but it cannot be true
for those at greater distances that are within the same age range.
The overall picture favors the \citet{HWB06} claim also supported by
\citet{SCH+12} of a dynamical evolution
of the whole cloud with a very minor triggering effect, if any.

Regarding the youngest population of the clusters, there is no evidence that
any of them has stopped forming stars. This is confirmed by {\em Herschel}
observations that discovered protostars and protostellar clumps
\citep{HMB+10,DSM+10} in all the clusters they have mapped.
NGC~2244, REFL~09, and CMFT~10 would be more likely to be about to stop their
activity. That is expected for NGC~2244 at the center of the H{\sc II} region
because most of its surrounding interstellar material has been swept up.
The fraction of Class~I sources should bring additional evidence. 
In Sect.~\ref{s.ukidss} we have estimated the total number of members in 
clusters with $K_s<15.5$~mag. At this luminosity level, all Class~I sources
are easily detected in the four WISE channels since they have a positive
spectral index, making them significantly brighter than the completeness limit.
For instance, the expected WISE magnitudes for an object with $K_s=15.5$~mag
and a spectral index of 0 are 14.2, 13.2, 10.3, and 8.2~mag.
It is therefore meaningful to derive the fraction of Class~I objects by
comparing the total number of members from Table~\ref{t.stat} with the
number of Class~I indicated in Fig.~\ref{f.alpha}.
The smallest fraction is obtained for NGC~2244 with ~3\%, confirming that 
the activity of this cluster has severely declined. The highest fraction
of Class~I objects is found in PL~01 and PL~02 with 10-30\% and 10-20\%,
respectively. The wide range results from the uncertainty on the amount
of extinction between the clusters and us. This high fraction does not
prove triggering has occurred, but it certainly indicates that the OB star
influence is not inhibiting the star formation in these two clusters,
although their projected distance to NGC~2244 is only about 14~pc.

Our results need to be compared with other regions where the question of
the influence of the massive stars on the surrounding interstellar medium
is raised. For instance, \citet{RPRG13} analyzed {\em Herschel} data in the
\object{Carina nebula} complex and claim that triggered star formation by
radiative cloud compression is observed.
\citet{DZC05} searched for star forming regions triggered by a 
{\em collect and collapse} process that corresponds to the expansion of
the H{\sc ii} nebula. They confirm the role of this process for regions
such as \object{RCW~79} and \object{Sh2-212} in companion papers 
\citep{ZDC+06,DLK+08}. Regarding the RMC we do not exclude that this process
has contributed in the close interaction zone which includes NGC~2237, PL~01,
and PL~02. \citet{DO13} studied
another mid--infrared bubble, \object{[CPA2006] N14}, and concluded that this
effect was not consistent with the dynamical time scales and suggest that a
compression by a shock wave is more likely to have triggered the star
formation. \citet{TBP+13} studied the star formation in the infrared dark
cloud \object{SDC G18.928-0.031} associated with an H{\sc ii} region.
Despite the morphology of the complex, which suggests triggered star
formation, they did not find evidence that the massive clump is prone
to collapse because of the expanding H{\sc ii} region.
Although the general characteristics of these complexes, which associate a
molecular cloud with an H{\sc ii} region, look similar, their evolution varies.
This is likely the result of the 3D structures, which are generally not known.
In a projected image we may not always be sure which clump is exposed to the
increased radiation energy density with UV photons.
This is supported by the theoretical works of \citet{DB11,DB12}, which show
that star formation may occur in dense filaments protected from ionizing
sources, with pre-existing bubbles shaped by turbulence that mimics a
triggered star forming region geometry.

\begin{table*}
\caption{YSO candidates in the Rosette molecular cloud. The full table of 535
objects within 13 clusters is published electronically at the 
CDS (http://cdsweb.u-strasbg.fr/cgi-bin/qcat?J/A+A/), this subsample lists
only the candidates for the two new clusters CMFT~10 and CMFT~11.
The astrometry and $W$[1-4] photometry is from WISE; the $JHK_s$ is
from UKIDSS or 2MASS when $<12$~mag; $\alpha$ is the spectral index
measured from $K_s$ to $W3$.}\label{t.yso}
\begin{tabular}{ccccrccccccc}
\hline
\hline
WISE & GLON & GLAT & Cluster & $\alpha$\; & $J$ & $H$ & $K_s$ & W1 & W2 & W3 & W4 \\
\hline
J062744.87+053101.6& 205.315003& -2.726404&   CMFT10&  0.2&      & 17.56& 15.30& 13.44& 11.75&  9.62&  6.38 \\
J062753.87+053207.5& 205.316099& -2.684801&   CMFT10&  0.1& 18.63& 16.06& 13.98& 12.84& 11.26&  8.72&  6.51 \\
J062754.95+053159.7& 205.320100& -2.681903&   CMFT10&  0.7& 16.84& 15.13& 12.30& 10.40&  8.30&  5.83&  3.31 \\
J062736.15+053025.5& 205.307206& -2.763104&   CMFT10&  1.6&      &      & 17.50& 14.82& 13.78&  9.14&  6.36 \\
J062739.15+053105.5& 205.303103& -2.746899&   CMFT10&  0.6& 14.50& 12.59& 11.02&  9.25&  7.99&  4.70&  2.36 \\
J062754.48+053059.7& 205.334001& -2.691302&   CMFT10& -0.8& 15.88& 14.62& 13.69& 12.65& 11.95& 10.01&  8.09 \\
J062752.39+053043.7& 205.333892& -2.701098&   CMFT10&  0.5& 13.23& 12.78& 12.42& 11.73& 11.27&  6.52&  4.64 \\
J062752.01+053230.8& 205.306795& -2.688701&   CMFT10&  0.1& 17.35& 15.77& 14.74& 12.92& 12.01&  9.21&  5.80 \\
J062735.67+053132.9& 205.289595& -2.756196&   CMFT10& -0.2& 16.00& 14.95& 14.26& 13.31& 12.52&  9.49&  7.05 \\
J062735.81+053113.6& 205.294696& -2.758105&   CMFT10& -0.9& 16.01& 14.63& 13.86& 13.21& 12.59& 10.47&  7.25 \\
J062738.29+053130.0& 205.295396& -2.746903&   CMFT10&  0.4& 16.87& 15.01& 13.52& 11.60& 10.77&  7.51&  5.62 \\
J062739.48+053155.3& 205.291502& -2.739300&   CMFT10& -0.5& 15.28& 14.20& 13.48& 12.68& 12.05&  9.27&  7.08 \\
J062751.25+053253.8& 205.299694& -2.688498&   CMFT10& -0.2& 14.78& 13.55& 12.71& 10.88&  9.99&  7.77&  5.70 \\
J062753.31+053201.5& 205.316560& -2.687612&   CMFT10&  0.7& 18.92& 17.20& 15.87& 14.07& 12.43&  9.30&  6.61 \\
J062752.39+053245.6& 205.303939& -2.685325&   CMFT10& -0.3& 17.41& 14.97& 13.16& 11.77& 10.48&  8.48&  5.55 \\
J062734.88+053050.5& 205.298601& -2.764498&   CMFT10& -0.3& 13.67& 13.31& 13.25& 12.81& 12.82&  8.73&  6.33 \\
J063314.39+022917.4& 208.638310& -2.908001&   CMFT11& -0.1& 16.06& 14.62& 13.65& 12.56& 11.54&  8.63&  5.58 \\
J063316.15+023021.3& 208.625788& -2.893300&   CMFT11&  2.6&      &      & 16.38& 11.41&  7.87&  5.78&  0.99 \\
J063315.21+023014.2& 208.625800& -2.897699&   CMFT11&  1.3& 17.63& 15.70& 13.32& 11.55&  9.43&  5.80&  2.95 \\
J063318.71+023102.5& 208.620502& -2.878601&   CMFT11&  1.6&      & 19.16& 16.53& 13.34& 11.04&  8.09&  5.13 \\
J063320.25+023054.5& 208.625505& -2.873899&   CMFT11&  0.5& 18.30& 16.04& 13.75& 11.84&  9.98&  7.55&  4.57 \\
J063313.67+023016.2& 208.622295& -2.903097&   CMFT11&  0.2& 17.61& 15.16& 13.33& 11.88& 10.24&  7.79&  4.85 \\
J063315.19+023030.0& 208.621897& -2.895701&   CMFT11& -0.0& 16.10& 15.16& 14.58& 13.96& 12.84&  9.65&       \\
J063316.35+023012.1& 208.628490& -2.893699&   CMFT11& -0.6& 14.38& 13.39& 12.90& 11.55& 11.07&  8.80&       \\
J063319.63+023121.6& 208.617644& -2.872707&   CMFT11&  0.8&      & 18.46& 16.62& 14.38& 12.56&  9.79&  5.90 \\
J063316.86+023051.1& 208.619872& -2.886830&   CMFT11&  0.3&      & 17.20& 15.14& 14.02& 12.33&  9.59&  5.90 \\
J063316.22+023037.2& 208.622069& -2.890993&   CMFT11& -0.1& 17.43& 15.26& 13.92& 12.64& 10.92&  9.00&  4.72 \\
\hline
\end{tabular}
\end{table*}

\section{Conclusion}\label{s.conclusion}
We performed a large scale study of the Rosette complex with near and
mid--infrared data, which allowed us to map the extinction and the young
star cluster surface densities. We proposed a new method based on the
joint analysis of star color and density to identify embedded star
clusters and to derive the main characteristics of their YSO population.
In particular, we showed that extinction mapping is essential for removing
the fluctuations of the field star surface density resulting from the
interstellar medium environment.
Extinction is also critical to estimating the number of stars in excess,
i.e. YSOs, within a molecular cloud. 
We found 13 clusters in the RMC, two of them new discoveries. The total
number of YSOs within these clusters is between $4\,000$ and
$8\,000$ stars depending on the cluster locations along the line of
sight with respect to the parent molecular cloud. The luminosity function
of the YSOs peaks at $K_s\approx 14$~mag, and very few objects are beyond 
17~mag at $K_s$, which points out that UKIDSS is sensitive enough to reach
completeness in this region. 

We identified 535 individual members within the clusters using WISE
photometry. The analysis of their SED from 1 to 22~$\mu$m allowed us to
estimate the age and the current status of each cluster. We concluded that
they all have recent activity and that a triggering scenario for the star
formation in the Rosette is not consistent with all the observed cluster
ages. Despite the presence of OB stars in the \object{Rosette nebula}
center, no evidence of triggering or inhibiting star formation has been
found except for NGC~2244.

More generally, the method we have developed to investigate embedded clusters
can be applied to other active regions to better constrain the impact
of triggering in star formation. This process is mostly invoked on a simple
geometrical basis with little physical analysis since it provides a plausible
explanation of the presence of clusters close to an OB association. We
suggest that this assumption is actually too naive and that a deeper analysis, 
such as this one, is required to confirm the actual role of massive stars
on the surrounding star formation.

\begin{acknowledgements}
L.~Cambr\'esy wish to thank O.~Roos for her contribution in initiating
the project.
G.~Marton has been supported by the Hungarian Research Fund (OTKA K-104607)
and by the PECS program of the European Space Agency and the Hungarian Space
Office (PECS-98073).
This research was funded partly by the grant OTKA 101393.
This work is based in part on data obtained as part of the UKIRT Infrared
Deep Sky Survey.
This publication makes use of data products from the Two Micron All Sky
Survey, which is a joint project of the University of Massachusetts and
the Infrared Processing and Analysis Center/California Institute of
Technology, funded by the National Aeronautics and Space Administration
and the National Science Foundation.
This publication makes use of data products from the Wide-field Infrared
Survey Explorer, which is a joint project of the University of California,
Los Angeles, and the Jet Propulsion Laboratory/California Institute of
Technology, funded by the National Aeronautics and Space Administration.
\end{acknowledgements}

\bibliographystyle{aa}
\bibliography{biblio}

\begin{appendix}
\section{Comparison with the column density obtained from {\em Herschel} data}\label{app1}
The extinction map derived in this work relies on the reddening of background
sources detected at $H$ and $K_s$ by UKIDSS. A totally independent estimation
of the column density is proposed by \citet{SCH+12} from dust emission.
Their map is based on the gray-body fitting from 160 to 500~$\mu$m of 
the 1.8~deg$^2$  observed by the {\em Herschel} key program HOBYS
\citep{MZB+10}.
Using our extinction map as reference, we convert the {\em Herschel} column
density (noted $CD$ and expressed in cm$^{-2}$) into extinction as 
$A_V(FIR)=(CD-3.71\times10^{20})/5.11\times10^{20}$. 
This relation depends mainly on the dust opacity at all wavelengths, and
the gas-to-dust ratio through the value of $N_H/A_V$ that is supposed
constant. However, the situation is more complicated, as pointed out by a
recent study by \citet{RMP+13}, which discusses the variation in the dust
opacity with column density in Orion. They conclude that no single opacity
can be applied to the whole cloud, which has quantitative implications for 
interpreting of the column density from {\em Herschel} data. Differences
between the two column density maps are therefore expected.

\begin{figure}
\includegraphics[width=8.8cm]{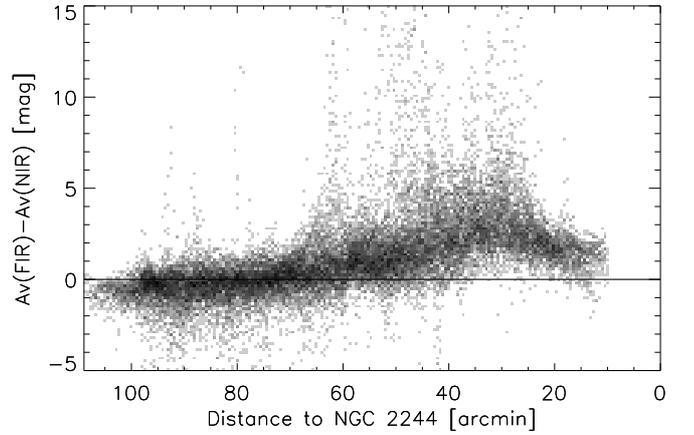}
\caption{Difference between the extinction obtained from dust emission in the
FIR and from dust reddening in the NIR versus the distance to NGC 2244. The
x-axis corresponds roughly to the galactic longitude and is reversed to match
the map orientation with east pointing left.}
\label{f.herschel1}
\end{figure}
\begin{figure}
\includegraphics[width=8.4cm]{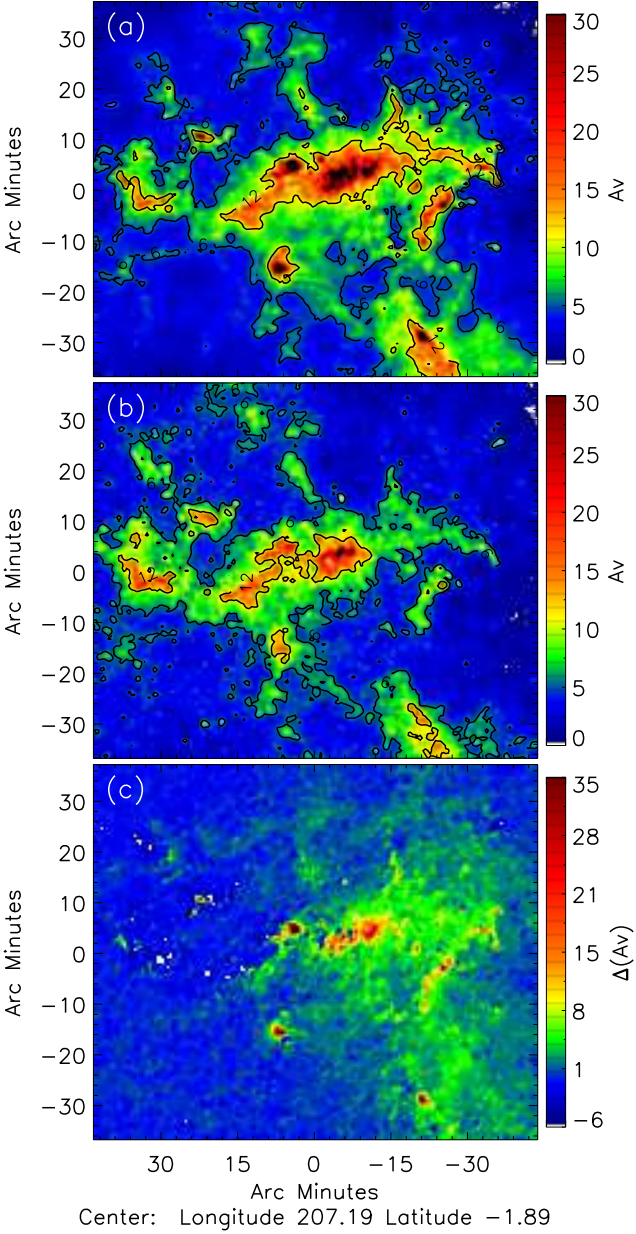}
\caption{(a) Extinction derived from dust emission using {\em Herschel} data.
(b) Extinction derived from UKIDSS $H-K_s$ color excess. 
(c) Difference between the two extinction map (a)-(b).
The color-map cuts are identical for (a) and (b) and set to display the
visual extinction from 0 to 30~mag. Contours for $A_V=6$ and 12~mag are
overlaid.}
\label{f.herschel2}
\end{figure}
\begin{figure}
\includegraphics[width=8.8cm]{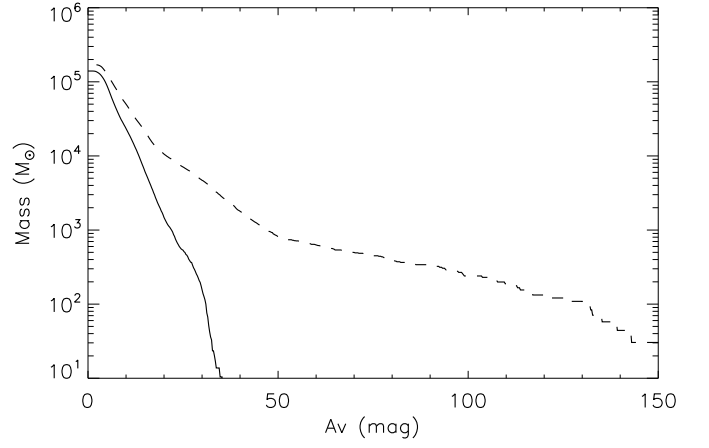}
\caption{Cumulative mass distribution from {\em Herschel} dust emission
(dashed line) and UKIDSS reddening (plain line) for the same field of view.}
\label{f.herschel3}
\end{figure}

Figure~\ref{f.herschel1} presents the difference between the two extinctions
as a function of the distance to NGC~2244, i.e. the OB stars. The excess of
extinction for the region at less than $\sim$50\arcmin\ from the NGC~2244
suggests that the optical depth derived from dust emission is overestimated near
the massive star cluster. The spatial distribution of the
excess is shown better in Fig.~\ref{f.herschel2}, which presents the two
extinction maps and their difference. The diffuse excess, which appears in
green in Fig.~\ref{f.herschel2}c recalls the ionization front of
Fig.~\ref{f.clusters}. The extinction excess likely results from the
emission of warmer grains in the diffuse envelop surrounding the dense
molecular cloud. 
A similar effect has been observed by \citet{RSA+13} in \object{NGC~6334}
when excluding the 70~$\mu$m to derive the column density. In addition, the
two maps exhibit strong differences of several tens of magnitudes for
limited areas that happen to match the star clusters. It suggests local
heating by YSOs, although it is true that real dense cores with
$A_V>150$~mag would not be revealed using $H-K_s$ since no background
star could be detected through such a column density.

Despite the obvious differences between the two extinction maps, it is
worth noting the total mass estimates shown in Fig.~\ref{f.herschel3} are
consistent. We obtained $1.7\times10^5$ and $1.4\times10^5$~M$_\odot$ for the
{\em Herschel} and the UKIDSS-based maps, respectively. The high extinctions
make only a very limited contribution to the total mass estimate because
they correspond to a small surface area.
Our analysis for the RMC confirms that {\em Herschel} column density maps must
be used with caution. Further comparisons between column density maps
derived with {\em Herschel} and extinction methods are required to better
understand the impact of parameters, such as the opacity (i.e. dust
emissivity), variations, and the presence of several dust components
along the line of sight.

\end{appendix}

\end{document}